# The Ferris ferromagnetic resonance technique: principles and applications

Amit Rothschild[*], Benjamin Assouline, Nadav Am Shalom, Nirel Bernstein, Goni Daniel, Gil Cohen, and Amir Capua[*]

Department of Applied Physics, The Hebrew University of Jerusalem, Jerusalem 9190401, Israel

The author to whom correspondence may be addressed: amir.capua@mail.huji.ac.il

**Abstract:**

Measurements of ferromagnetic resonance (FMR) are pivotal to modern magnetism and spintronics. Recently, we reported on the *Ferris FMR* technique, which relies on large-amplitude modulation of the externally applied magnetic field. It was shown to benefit from high sensitivity while being broadband. The Ferris FMR also expanded the resonance linewidth such that the sensitivity to spin currents was enhanced as well. Eventually, the spin Hall angle ($\theta_{SH}$) was measurable even in wafer-level measurements that require low current densities to reduce the Joule heating. Despite the various advantages, analysis of the Ferris FMR response is limited to numerical modeling where the linewidth depends on multiple factors such as the field modulation profile and the magnetization saturation. Here, we describe in detail the basic principles of operation of the Ferris FMR and discuss its applicability and engineering considerations. We demonstrated these principles in a measurement of the orbital Hall effect taking place in Cu, using an Au layer as the orbital to spin current converter. This illustrates the potential of the Ferris FMR for the future development of spintronics technology.

Spintronics is a promising candidate for future nonvolatile electronics. It is expected to complement CMOS technology and deliver improvements in power, density, and speed. In recent years, studies of magnetization dynamics have proven pivotal to the rapid evolution of spintronics technology. The ferromagnetic resonance (FMR) technique is a key method for this purpose. It can be used to resolve basic material properties including the magnetization saturation ($M_s$), Gilbert damping ($\alpha$), and for discovering new material systems and physical interactions such as the charge to spin conversion efficiency by the spin Hall effect (SHE).

Classical FMR experiments are generally implemented in two ways: 1) The cavity FMR [1,2] is narrowband but offers very high sensitivity down to the single atomic layer. 2) The stripline FMR [3,4] is broadband but generally has lower sensitivity as compared to the cavity FMR due to reduced RF field confinement. Recently, we demonstrated an FMR technique that is both broadband and sensitive [5]. Instead of utilizing a small-amplitude modulation of the externally applied magnetic field [1-4], a large-amplitude modulation was applied. This resulted in a proportionally large signal such that the sensitivity was enhanced by nearly two orders of magnitude without compromising on the bandwidth. The modulation was achieved by placing permanent magnets on a spinning disk, hence the technique was referred to as the *Ferris FMR*. Most importantly, the large-amplitude modulation turned out to expand the linewidth. Consequently, the conversion of the SHE, known as the spin Hall angle, $\theta_{SH}$, was resolvable at low current density while also reducing the Joule heating, thus making the wafer-level measurements of $\theta_{SH}$ possible [5].

Despite the advantages of the large-amplitude modulation, the Ferris FMR has one main drawback: it is limited to a numerical analysis apart from a few special cases that are less relevant experimentally. Additionally, the measured signal also depends on the transient absorption profile thus making it less intuitive. Nevertheless, the Ferris FMR offers substantial design flexibility since the absorption spectra depend on the geometry of the rotating disc and magnets. For example, by proper choice of the number of magnets, their spacing, and physical dimensions, different modulation profiles can be achieved to a meet a desired requirement such as enhanced sensitivity to spin currents or improved signal to noise ratio (SNR).

The aims of this publication are to elaborate on the principles of the Ferris FMR as well as to present advanced engineering concepts. We start by describing the operation of the Ferris FMR and analyze a typical measurement. Then we discuss the role of the modulation profile and the relevant engineering considerations. Finally, we present a measurement of spin currents generated by the orbital Hall effect (OHE) taking place in Cu, using an Au orbital to spin current conversion layer.

The Ferris FMR setup is presented in Fig. 1. A magnetic sample is placed on a stripline waveguide. The externally applied time-dependent magnetic field, $H(t)$, is generated by placing permanent magnets on a spinning disc. $H(t)$ can be designed to be either out-of-plane or in-plane. Here we implement the in-plane configuration, which is required for measurements of spin currents. The magnitude of $H(t)$ is controlled by varying the sample-magnet distance, $l$, which is achieved by placing the disc on a translation stage. The absorption spectrum is measured by scanning, $l$, while recording the power of the RF signal at the output of the waveguide using an RF diode and a lock-in amplifier. Here, a disc consisting of four magnets was used and the in-plane $H(t)$ was achieved by placing pairs of magnets arranged in opposite polarity. Each magnet was $10 \times 17\ mm^2$ while the disc was $13\ cm$ in diameter. For a given $l$, we indicate the maximal value of $H(t)$ by $H_0$. The minimal incremental step of the translation stage was $0.05\ \mu m$ so that $H_0$ was controllable to an accuracy of $0.6\ \mu T$ at the closer end, well beyond any requirement of an FMR experiment. The maximal value of $H_0$ was $\sim 0.46\ T$.

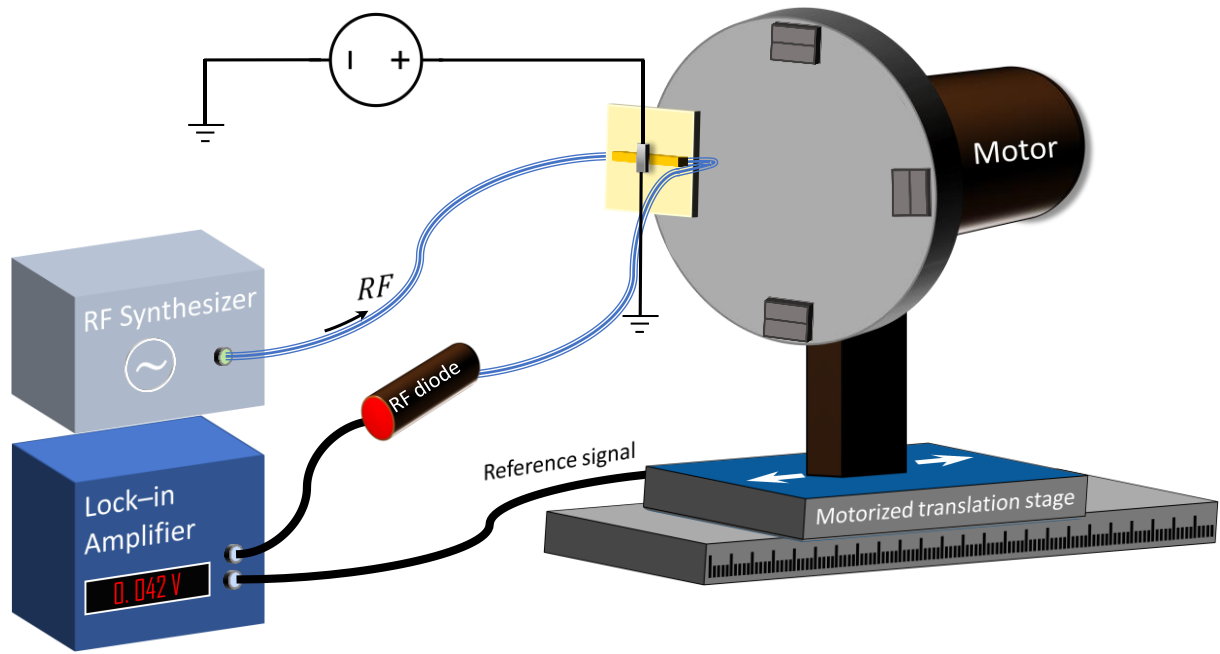

**Figure 1. Ferris FMR setup. The time-varying magnetic field $H(t)$ is generated by rotating the magnetic disc. The RF signal is passed through the waveguide and is detected using an RF diode and a lock-in amplifier.**

To explain the operation of the Ferris FMR, we examined the temporal profiles of the absorption and the applied field, as well as the action of the lock-in amplifier. An example of a measured $H(t)$ profile is presented in Fig. 2(a) for $l = 7.67\ mm$, indicated by the open blue circles. To a good agreement, $H(t)$ can be fitted to a Gaussian function presented by the black solid line. Figure 2(b) presents $H(t)$ for a full range of $l$, readily showing that the Gaussian profile persists as $l$ is scanned. As an example, in Fig. 2 we marked the resonance field, $H_{res}$, by the vertical black dashed line that corresponds to an RF frequency $f$. The difference $|H(t) - H_{res}|$ determines the momentary absorption, $A(t)$. As long as $H_0 < H_{res}$, $A(t)$ exhibits a single absorption peak per cycle. When $H_0 > H_{res}$, $H(t)$ crosses $H_{res}$ twice per cycle leading to a double-peaked $A(t)$. The voltage drop on the RF diode can be described by $V_{RF}(t) = V_0 \cdot (1 - A(t))$, where $V_0$ is the zero-absorption voltage. Figure 2(c) presents a numerical calculation of $V_{RF}(t)$ in which $H(t)$ was modeled as a Gaussian function $G(t) = H_0 \cdot e^{-\frac{1}{2}\frac{t^2}{\sigma^2}}$, where $\sigma$ is the standard deviation. The signal is recorded on a lock-in amplifier, therefore $V_0$ can be neglected and thus also the polarity of $V_{RF}(t)$. $V_{RF}(t)$ was calculated assuming a Lorentzian absorption line leading to $V_{RF}(t) = 1/(1+\beta)^2$ with $\beta = \gamma(H(t) - H_{res})/2\pi\alpha f$, where $\gamma$ is the gyromagnetic ratio. Figure 2(c) presents three cases in which $H_0$ is smaller, equal, and greater than $H_{res}$ illustrating the evolution from a single-peaked absorption profile to the double-peaked profile. The calculation was carried out at 5 $GHz$ and $\alpha = 0.01$. $H_{res}$ was determined by Kittel's formula, $f = \mu_0 \frac{\gamma}{2\pi}\sqrt{H_{res}(H_{res} + M_s)}$, assuming $M_s = 8 \times 10^5 \, A \cdot m^{-1}$ and $\sigma = 358.4 \, \mu sec$. The signal recorded on the lock-in amplifier, $V_{LIA}$, is determined by projecting $V_{RF}$ on the fundamental harmonic. To that end, $H(t)$ was modeled by $H(t) = \sum_i G(t - i/f_{mod})$, where $f_{mod}$ is the modulation frequency. Figure 2(d) presents the calculated $V_{LIA}(H_0)$ together with the corresponding Lorentzian absorption function $L(H_0)$. Instead of measuring the differential absorption as in small-amplitude modulation-based techniques, $V_{LIA}(H_0)$ resembles more closely the actual absorption lineshape which is asymmetric, and much broader as compared to $L(H_0)$. Moreover, the peak of $V_{LIA}(H_0)$ occurs at a slightly higher $H_0$ value than $H_{res}$. We denote this field by $H_{peak}$. These two effects stem from the fact that the resonance conditions are met even when $H_0 > H_{res}$. In contrast to conventional FMR methods, in the Ferris FMR there is some residual absorption even when $H_0$ is increased well beyond $H_{res}$. The higher $H_0$ is, the less time the residual absorption takes place and eventually $V_{LIA}(H_0)$ decreases. Accordingly, for an ideal on-off modulation profile, the transition of $H(t)$ is instantaneous and the residual absorption vanishes. This calculation is presented as well in Fig. 2(d) by the solid red line which accurately reproduces $L(H_0)$. This result illustrates that the larger modulation associated with the larger $H_0$ does not affect the lineshape since the RF power is constant for all $H_0$ and only the absorption by the sample is modulated.

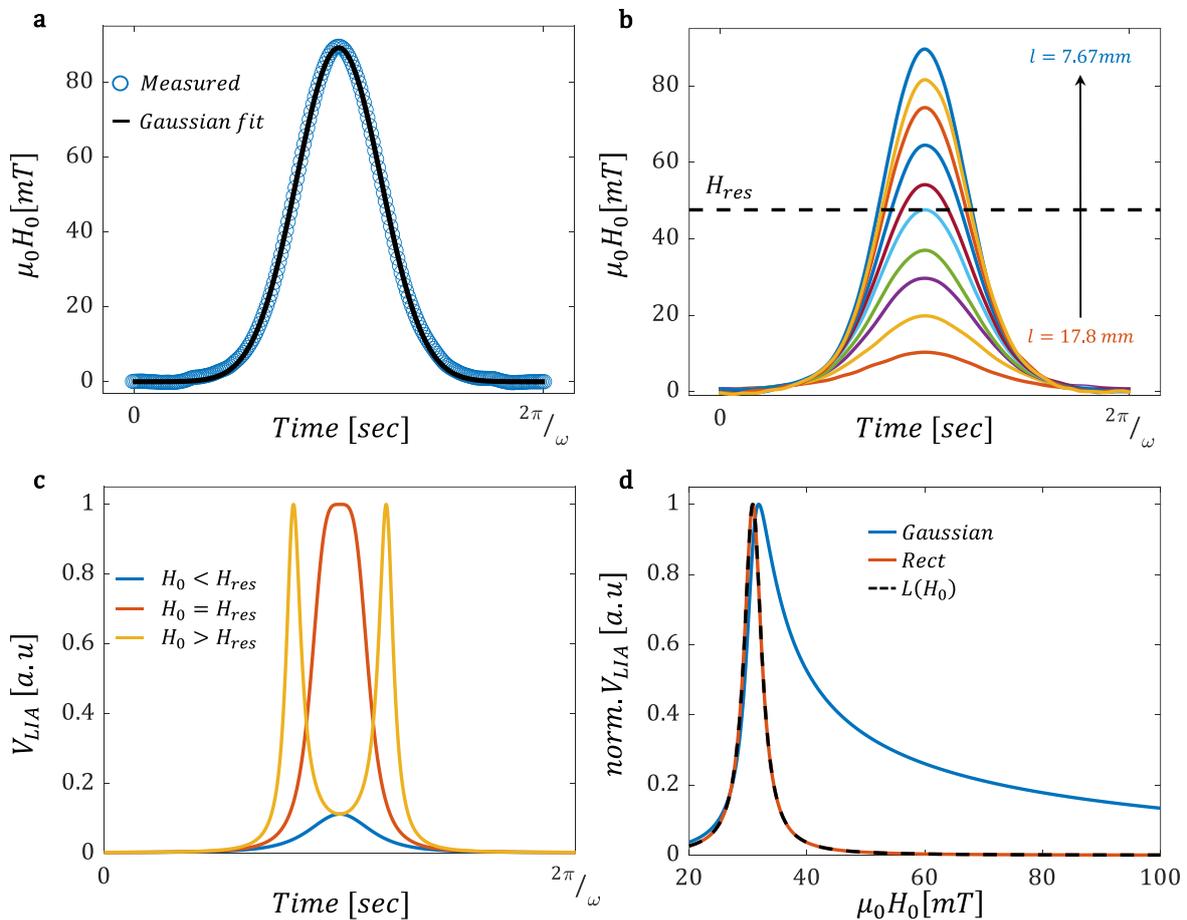

**Figure 2. Large-amplitude modulation and construction of the Ferris FMR trace. (a) Measured $H(t)$ (blue open circles) at $l = 7.67 \, mm$ along with the Gaussian fit (black solid line). (b) Measured $H(t)$ for various $l$ values. Horizonal dashed line indicates an example of an FMR absorption occurring at $H_{res}$. (c) $A(t)$ profiles for $H_0$ below (blue), on (red), and above (yellow) resonance. (d) Calculation of the Ferris FMR lineshapes of a Gaussian (blue) and a rectangular (red) modulation with $L(H_0)$ (dashed black). The simulations in (c) and (d) were carried out at $f = 5 \, GHz$, $M_s = 8 \times 10^5 \, A/m$, and $\alpha = 0.01$.**

Ferris FMR measurements were carried out on a sample of 30 Cu| 1 Au | 7.5 Py | 2.5 TaN (numbers indicate layer thickness in nm). The sample was grown on a Si/SiO2 substrate by Ar-based DC magnetron sputtering. The measurements were carried out with $l$ typically in the range of $7 - 24\ mm$, while $f_{mod}$ was 300 $Hz$. The sample under test was $0.7 \times 0.5\ cm^2$ in size. Homogeneity of $H(t)$ was verified by testing smaller samples of ~$0.5 \times 0.5\ mm^2$ for which the FMR response remained unchanged. The measured FMR spectra are presented in Fig. 3(a). The predicted asymmetric lineshapes are readily seen. However, the lineshapes at $f > \sim 4\ GHz$ broaden towards $H > H_{res}$ as calculated, while at lower $f$ the broadening occurs towards $H < H_{res}$. The deviation from the predicted lineshapes at low $f$ stems from a slight misalignment between the easy magnetization axis of the square dies and the externally

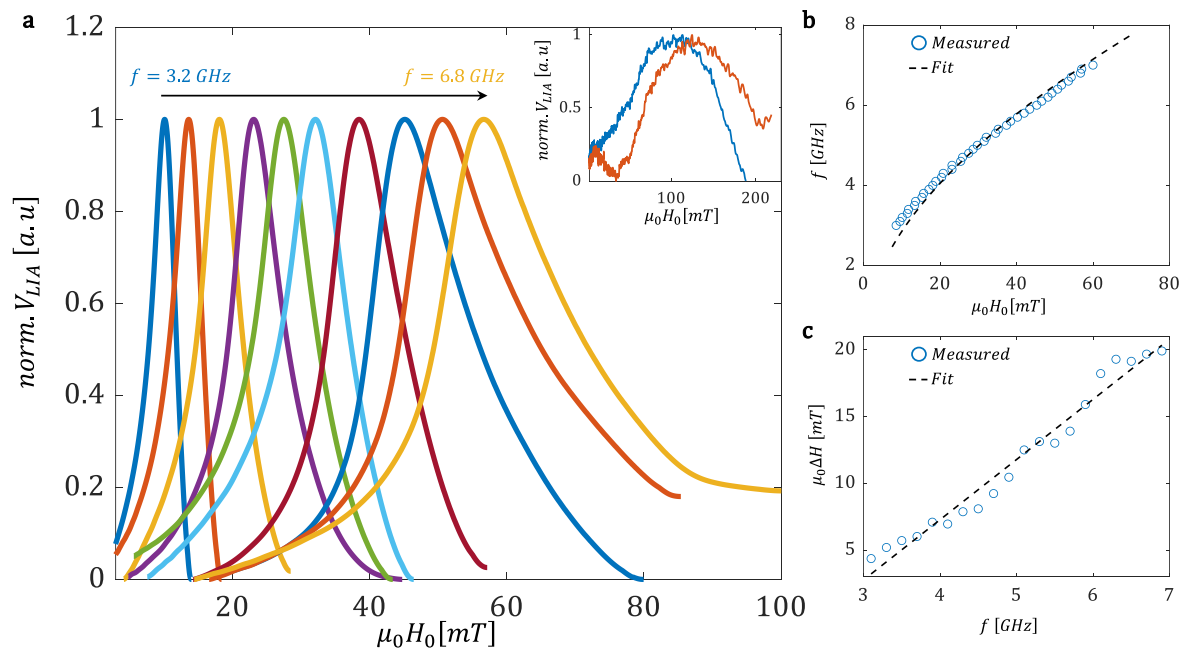

**Figure 3. Ferris FMR measurements of the 30 Cu|1 Au| 7.5 Py sample (thicknesses in $nm$). (a) Measured Ferris FMR traces. Traces are normalized to unity at the peak. Inset presents measurement of a $1.5\ nm$ of Py at 3 (blue) and $3.2$ (red) $GHz$. (b) Extracted $f$ vs. $H_{res}$ (blue open circles) and the Kittel formula (dashed black line). (c) Measured $\Delta H_F$ as a function of $f$ (blue open circles) and a linear fit (black dashed line).**

applied field which was not modeled. The traces also demonstrate high SNR of the Ferris FMR technique. From the root mean square error, the average SNR was found to be $\sim 5000$ for which a detection limit of $6.9 \times 10^{10}\ \mu_B$ was extracted representing an improvement by approximately two orders of magnitude as compared to conventional FMR methods [4]. The inset in Fig. 3(a) presents the absorption line of a Py sample of 15 Å, illustrating the ability to measure much thinner magnetic samples. From these measurements, the $f$ vs. $H_{res}$ dispersion relation was constructed, as presented in Fig. 3(b). Here we applied the approximation $H_{res} \cong H_{peak}$, which was validated numerically and found to lead to a typical error of $\pm 0.5\ mT$. From $f(H_{res})$, $M_s$ was determined to be $8.15 \times 10^5 \pm 415\ A \cdot m^{-1}$, agreeing well with known values for Py [6]. A slight deviation from Kittel's formula is seen at frequencies below $\sim 4\ GHz$ illustrating once more the contribution of the misalignment between the easy magnetization axis of the square dies and the externally applied field. The frequency dependence of the Ferris FMR Linewidth, $\Delta H_F$, defined by the full width at half maximum, is presented in Fig. 3(c). It is inherently broader than the linewidth of $L(H_0)$, $\Delta H_L$, that is given by $\Delta H_L = 4\pi\alpha f/\gamma$. Accordingly, we define $\Delta H_F = A_c \cdot \Delta H_L$ where $A_c$ is the expansion factor. Currently, the most reliable method to calibrate $A_c$ for a given implantation, is by comparison to a reference conventional FMR method. In our case we used an optically probed spin-torque FMR (OSTFMR) [7,8]. From comparisons of measurements carried out by a reference FMR method, $A_c \sim 9$ was extracted. However, in our simulations we found $A_c = 3.65$, indicating the role of inhomogeneous broadening in our measurement [9].

The expansion of the Ferris FMR linewidth by $A_c$ is beneficial for measurements of the SHE and OHE in which the resonance linewidth is modulated by the anti-damping torque. In principle, it is expected that $\Delta H_F = A_c \cdot 4\pi\alpha f/\gamma$, however $A_c$ may also depend on $\alpha$, $M_s$, and $f$. This points to the need to examine the influence of the profile of $H(t)$, $\alpha$, $M_s$, and $f$ on $\Delta H_F$. To this end, we calculated the sensitivity of $\Delta H_F$ to variations in $M_s$ and $\alpha$ for a Gaussian, piecewise linear, and sinusoidal modulation profiles. The different transition rates, $\partial H(t)/\partial t$, of these modulations significantly affect the absorption profiles and the respective Ferris FMR lineshapes. Figures 4(a) – 4(c) present the case of the Gaussian $H(t)$. The absorption spectra for different values of $\alpha$ are presented in Fig. 4(b) and are indicated by the solid lines in addition to the corresponding $L(H_0)$ traces marked by the dashed lines. The calculation was carried out at $5\ GHz$ and $M_s = 8 \times 10^5\ A \cdot m^{-1}$. As compared to $\Delta H_L$, it is seen that the dependence of $\Delta H_F$ on $\alpha$ is more prominent as summarized in the inset. $\Delta H_F$ depends linearly on $\alpha$ resulting in $\partial \Delta H_F/\partial \alpha = 1136\ mT$ while $\partial \Delta H_L/\partial \alpha = 357\ mT$, namely, the application of the Gaussian profile enhances the sensitivity of the Ferris FMR to $\alpha$ by $\sim 3.2$. Figure 4(c) presents a similar analysis for $M_s$ taking $\alpha = 0.01$ while the inset presents $\Delta H_F$ and $\Delta H_L$ vs. $M_s$. Obviously, $\Delta H_L$ is not affected by $M_s$, however, it is seen that $\Delta H_F$ decreases with $M_s$. This implies that $\alpha$ can only be extracted numerically following extraction of $M_s$ by Kittel's formula. Figures 4(d) - 4(f) present the same analysis for the sinusoidal profile case. Figure 4(e) illustrates the $\alpha$ dependent responses resulting in $\partial \Delta H_F / \partial \alpha = 665\ mT$. In this case as well, the sinusoidal $H(t)$ results in a higher sensitivity to $\alpha$ as compared to conventional FMR measurements but to a lower degree as compared to the Gaussian profile. The influence of $M_s$ on the Ferris FMR spectra is summarized in Fig. 4(f). The inset again shows a dependence of $\Delta H_F$ on $M_s$ that is nonlinear in this case. As compared to the Gaussian profile, $M_s$ affects the linewidth more significantly. Finally, when a linear $H(t)$ profile is applied as presented in Figs. 4(g)-4(i), an inherently broader Ferris FMR trace results as compared to $L(H_0)$ but, at the same time, $\Delta H$ is less affected by $\alpha$ as readily seen in the inset. In this case $\partial \Delta H_F/\partial \alpha = 319\ mT$ and is even slightly smaller than the sensitivity of a conventional FMR measurement. In contrast, $M_s$ in this case significantly affects $\Delta H$ as seen in Fig. 4(i) where an exponential decay of $\Delta H$ with $M_s$ is seen. To quantify the effect of $f$ on $A_c$, we calculated $A_c$ across a range of frequencies shown in Fig. 4 (j). In the case of the Gaussian modulation, $A_c$ remains constant with respect to $f$. For a sinusoidal modulation, a slight dependence on $f$ is seen. Lastly, for the case of a piecewise linear modulation, a noticeable linear dependence with respect to $f$ is found. Overall, we conclude that the Gaussian profile results in the highest sensitivity to $\alpha$ and is least affected by $M_s$ and $f$.

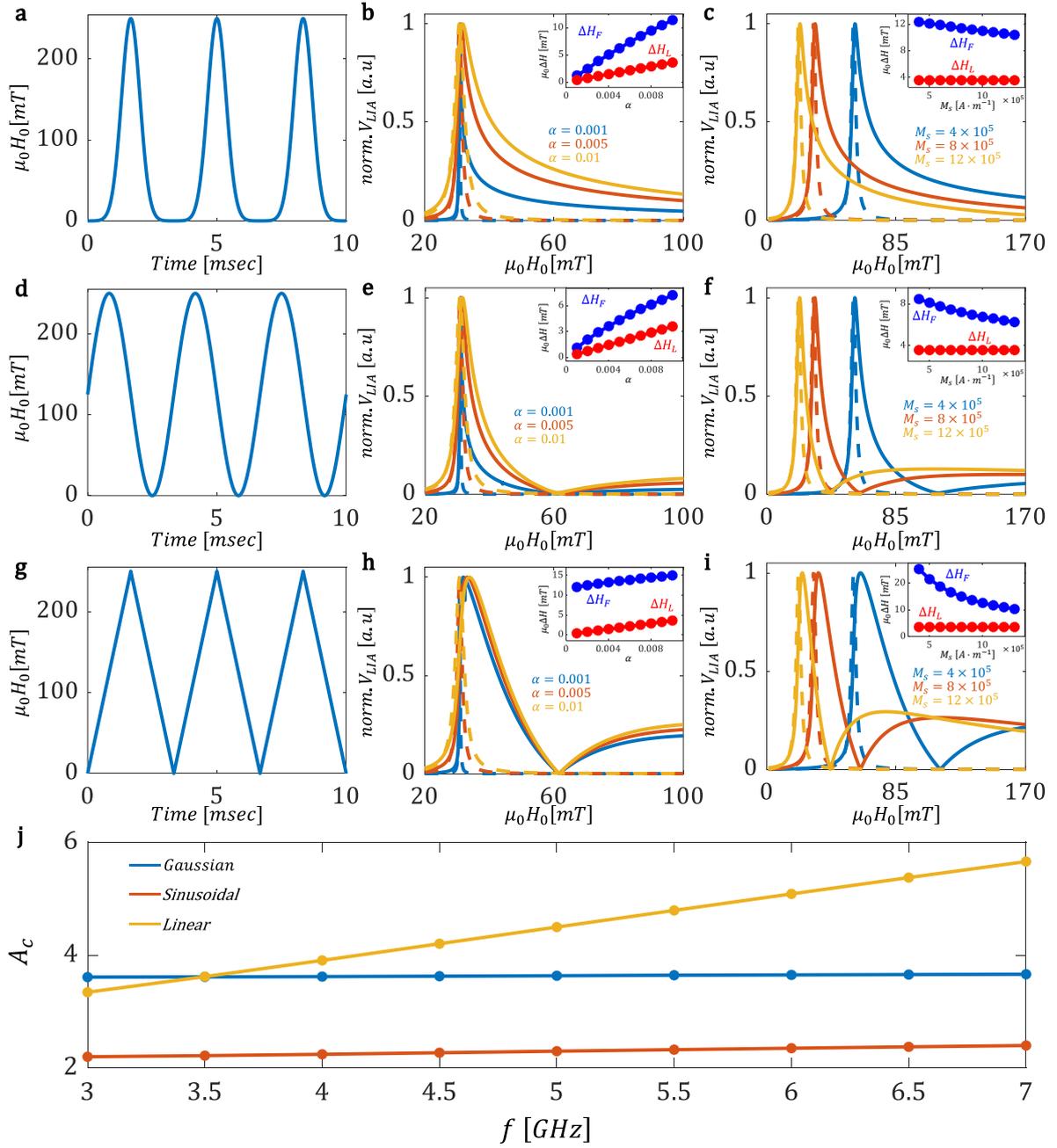

**Figure 4.** Calculation for different modulation profiles of $H(t)$. (a)-(c) calculation for a Gaussian profile. (a) $H(t)$. (b) Ferris FMR response at $5\ GHz$ for $\alpha = 0.001$ (solid blue), $0.005$ (solid red), and $0.01$ (solid yellow) and the corresponding $L(H_0)$ (dashed lines and same color code). The calculation was carried out for $M_s = 8 \times 10^5\ A \cdot m^{-1}$. Inset: $\Delta H_F$ (blue) and $\Delta H_L$ (red) as a function of $\alpha$. (c) Calculation of Ferris FMR response at $5\ GHz$ for $M_s = 4 \times 10^5$ (solid blue), $8 \times 10^5$ (solid red) and $12 \times 10^5$ (solid yellow) $A \cdot m^{-1}$ and the corresponding $L(H_0)$ (dashed lines, same color code). Calculation was carried out for $\alpha = 0.01$. Inset: $\Delta H_F$ (blue) and $\Delta H_L$ (red) vs. $M_s$. (d)-(f) Same as in (a)-(c), but for a sinusoidal $H(t)$. (g) - (i) Same as (a) - (c), but for a linear $H(t)$. (j) Calculation of $A_c$ as a function of $f$ for Gaussian (blue), Sinusoidal (red) and Linear (yellow) $H(t)$ profiles. Calculations were carried out using $M_s = 8 \times 10^5\ A/m$, and $\alpha = 0.01$.

    Finally, we demonstrated Ferris FMR measurements in the presence of spin currents. In our sample, orbital currents are generated in the Cu layer by the OHE [5] and the Au is responsible of converting them into detectable spin current by its spin orbit coupling (SOC). It is well known that Cu does not exhibit a large $\theta_{SH}$ [10], and even though SOC is large in Au, $\theta_{SH}$ of Au is negligible [11]. Namely, the spin currents in the trilayer emerge from the orbital currents generated by the Cu layer and their conversion by the Au layer as also demonstrated in Refs. [12-14]. Figure 5 presents the modulation of $\Delta H_F$ as a function of the applied charge current density, $J_c$. The measurements were carried out at $f = 5\ GHz$, using the Gaussian $H(t)$ for which $A_c$ is independent of $f$. We describe the charge to spin conversion efficiency in terms of an effective $\theta_{SH}$, $\theta_{SH}^{eff}$, that describes the Cu/Au bilayer. $\theta_{SH}^{eff}$ can be extracted using the following relation: $\Delta H_F \cong A_c \frac{\sqrt{H_{res}(H_{res}+M_S)}}{(1+\alpha)^2(H_{res}+0.5M_S)} \frac{\hbar}{eM_s t_{FM}} \cdot J_s$, where $\hbar$ is the reduced Planck's constant, $e$ is the electron charge, $t_{FM}$ is the thickness of the ferromagnetic layer, and $J_s$ is the spin current density injected into the Py and is given by $J_s = \theta_{SH}^{eff} \cdot J_c$ [5]. $J_c$ is the current density in the Cu layer. It is approximately the applied current passing through the entire device since the conductivity and thickness of the Cu layer are significantly higher as compared to the other layers. The sensitivity of the Ferris FMR to the effective $\alpha$ enables one to detect the generated spin currents at low $J_c$. Figure 5 presents the $\Delta H_F$ data as a function of $J_c$ as measured on the same die of $0.7 \times 0.5\ cm^2$ in size where a maximal $J_c$ of $3 \times 10^9\ A \cdot m^{-2}$ was applied. From the measurement, we find $\theta_{SH}^{eff} \sim 0.04$. To extract the intrinsic orbital Hall conductivity of Cu, a systematic study of the thickness dependence of $\theta_{SH}^{eff}$ is required and is beyond the scope of the present work.

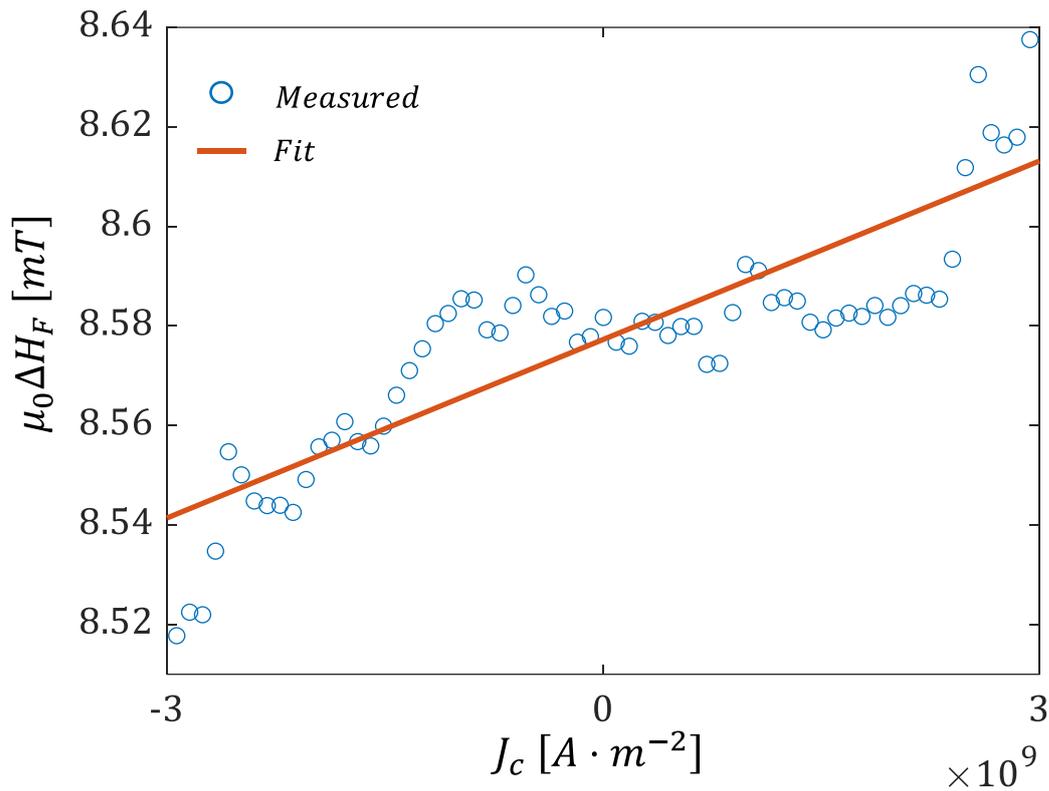

**Figure 5. $\Delta H_F$ vs. $J_c$. Open blue circles indicate the measured $\Delta H_F$ and the red solid line is the linear fit**

In conclusion, the Ferris FMR is a fundamental addition to the family of FMR techniques. Here, we reviewed its basic principles of operation and discussed its applicability and engineering considerations. Despite benefiting primarily from very high sensitivity to the FMR signal, its limitation arises from the fact that a numerical analysis is required to complement the measurements. This work is a first step in the avenue of large-amplitude modulation-based FMR experiments and has much room for further design and engineering.

## Acknowledgments

A.C. acknowledges the support from the Israel Science Foundation (Grant No. 1217/21), the Peter Brojde Center for Innovative Engineering and Computer Science, and from the Center for Nanoscience and Nanotechnology of the Hebrew University of Jerusalem.

## Data availability

The data that support the findings of this study are available from the corresponding author upon reasonable request.